\documentstyle[12pt]{article}
\def\be{\begin{equation}}
\def\ee{\end{equation}}
\def\bea{\begin{eqnarray}}
\def\eea{\end{eqnarray}}

\begin{document}
\begin{center}
{\Large{\bf Non-commutative D$p$-Brane in General Background Fields}}

\vskip .5cm
{\large Maryam Zoghi and Davoud Kamani}
\vskip .1cm
{\it Faculty of Physics, Amirkabir University of Technology (Tehran Polytechnic)\\
P.O.Box: 15875-4413, Tehran, Iran}\\
{\it e-mails: zoghi , kamani@aut.ac.ir}\\
\end{center}

\begin{abstract}

We investigate non-commutativity of open strings,
attached to a D$p$-brane, in the presence of the
linear dilaton, tachyon, $U(1)$ gauge field
as well as constant anti-symmetric $B$-field backgrounds.
Non-commutativity parameter, open string metric
and some special cases will be studied. Mode-dependent
non-commutativity, inspired by the tachyon field,
will be discussed in detail.

\end{abstract}
\vskip .5cm



\newpage
\section{Introduction}

D-branes, as inevitable objects in string theory \cite{1},
have been studied from various points of view.
On the other hand we have non-commutative geometry which has
been used as a framework for many subjects in physics and chiefly
in high energy phsyics.
Advent of non-commutative string theory in the $B$-field
background and appearance of non-trivial commutators in
spatial directions of D-brane \cite{2,3,4,5}, revived
investigation of non-commutativity in the string
theory with different backgrounds fields like
dilaton field \cite{6,7}, tachyon field \cite{8,9}, as well as
moving D-branes.

Unstable D-branes in background fields are mainly studied in
world-volume field theory and boundary state descriptions
\cite{10, 11, 12}. In this paper we consider a D$p$-brane
of the bosonic string theory in the presence of a $U(1)$
gauge field, linear dilaton, tachyon and Kalb-Ramond $B$-field,
simultaneously.  This is a system with generalized
set-up. For this system we obtain the boundary conditions
and propagator of an open string, attached to the D$p$-brane. This
propagator enables us to extract a set of open string metrics and
non-commutativity parameters. We observe that these
variables depend on the open string modes. On these variables
some scaling limits and approximations will be investigated.
Finally, we evaluate our achievements for the D1-brane and
D2-brane as special cases.
\section{Extended boundary conditions of open string}

We consider a D$p$-brane in a general
background, which contains the anti-symmetric field $B_{\mu \nu}$,
the dilaton field $\Phi$, the open string tachyon field $T(x)$ and
a $U(1)$ gauge field $A_\alpha$
which lives on the D$p$-brane world-volume.
We shall use the indices $\alpha\in\{0,1,\cdot\cdot\cdot,p\}$
and $i \in \{p+1,\cdot\cdot\cdot,d-1\}$ for directions along the
world-volume of the D$p$-brane and perpendicular to it, respectively.
Therefore, the string sigma model action in these background fields
has the following feature
\bea
S &=&
\frac{1}{4\pi\alpha'} \int_\Sigma d^2\sigma ({\sqrt{-h}} h^{ab} g_{\mu
\nu}\partial_a X^\mu \partial_b X^\nu - 2\pi i \alpha'
\epsilon^{ab}B_{\mu \nu}\partial_a X^\mu \partial_b X^\nu +
\alpha'\sqrt{-h} \Phi R^{(2)})
\nonumber\\
&+& \frac {1}{2 \pi \alpha' } \int_{\partial\Sigma}
d\tau \{-2\pi i \alpha' A_\alpha
\partial_\tau X^\alpha + T(X)\},
\eea where $\Sigma$ indicates the string worldsheet which has the
boundary $\partial \Sigma$ and the metric $h_{ab}$ with $h =
-\det h_{ab}$. The scalar curvature $R^{(2)}$ is constructed from
the metric $h_{ab}$. The space-time metric also is $g_{\mu \nu}$.
Note that the indexes of worldsheet and spacetime take their
values from the sets $a,b \in \{\sigma, \tau\}$ and $\mu,\nu \in
\{0,1, \cdot\cdot\cdot, d-1\}$, respectively.

The variety of the background fields enables us to restrict them to
obtain a solvable model. For this we take the background fields $g_{\mu \nu}$ and
$B_{\mu \nu}$ to be constant with $g_{\alpha i}=0$ and only
$B_{\alpha \beta}\neq 0$. In addition, for the $U(1)$ gauge
field we elect the gauge
$A_\alpha=-\frac{1}{2}F_{\alpha\beta}X^\beta$, where the field
strength $F_{\alpha \beta} =\partial_\alpha A_\beta - \partial_\beta A_\alpha$
is constant. Moreover, we suppose that the dilaton field
has a linear form along the brane world-volume,
i.e, $\Phi = a_\alpha X^\alpha $
where the parameters $\{a_\alpha |\alpha =0,1, \cdot \cdot \cdot ,p\}$
are constant. In the conventional case the dilaton is an arbitrary
function, which causes appearance of third order derivative in the
action, and hence is put away \cite{13}. In the
action (1) the scalar curvature $R^{(2)}$ contains second order
derivative. This implies that by considering a linear
dilaton only the usual second order derivative is obtained.

Remember that the exponential of the dilaton field gives the
strength of the string coupling. So the linear dilaton
background describes a world in which strings are weakly
coupled for large negative $x$ and strongly coupled for
large positive $x$. Thus, one could worry about the
reliability of the formalism in such set-up. However,
adding a tachyon background term of the form
$T_0 \exp(u\cdot X)$ suppresses the contribution of the strongly
coupled region, and this keeps things under control \cite{14}.
The conventional inhomogeneous tachyon profile is a linear
form which appears as a squared term in the boundary of the
string action \cite{15}. This motivates us to consider tachyon
profile as
$T(X)= \frac {1}{2} U_{\alpha\beta} X^\alpha X^\beta$ with
the constant symmetric matrix $U_{\alpha \beta}$,
which originates from the expansion of exponetial form
$T_0 \exp(u_\alpha X^\alpha )$ for small parameters
$\{u_\alpha |\alpha =0,1, \cdot \cdot \cdot ,p\}$.
This form of the tachyon field accompanied by the linear
dilaton gives a Gaussian theory, and hence is exactly solvable \cite{16}.

Reparametrization invariance of the bulk part of the
action (1) enables us to choose the conformal gauge for the worldsheet
metric, i.e, $h_{ab} (\sigma,\tau )= e^{\rho(\sigma ,\tau)} \eta_{ab}$.
Note that due to the presence of the dilaton, the action does not have
the Weyl symmetry. Thus, $\rho (\sigma , \tau)$ is a nonzero worldsheet field.

Vanishing the variation of the action leads to the equations of motion
for the worldsheet fields $X^\mu$ and $\rho$ as in the following
\bea
&~& \partial^2 X_\mu + \frac {1}{2} a_\mu \partial^2 \rho =0,\\
&~& a_\alpha \partial^2 X^\alpha = 0,
\eea
where $\partial^2 =\eta^{ab}\partial_a \partial_b$.
Now we consider a non-critical string theory, i.e.
$a^2=a_\alpha a^\alpha \propto d-26 \neq 0$. Then the Eqs. (2) and (3)
split into $\partial^2 X^\mu=\partial^2 \rho =0$.
In addition, vanishing of this variation defines the boundary
conditions of the string. For example, for the open string end
at $\sigma=0$ we receive the boundary equations
\bea
&~& (g_{\alpha \beta} \partial_\sigma X^{\beta} + 2\pi i
\alpha'{\cal{F}}_{\alpha \beta}
\partial_\tau X^{\beta} + U_{\alpha \beta} X^{\beta})_{\sigma=0} = 0,
\nonumber\\
&~& (\delta X^i)_{\sigma=0} = 0,
\nonumber\\
&~& (a_\alpha \partial_\sigma X^\alpha)_{\sigma=0} =0.
\eea
Here,
${\cal{F_{\alpha \beta}}} =B_{\alpha \beta} - F_{\alpha \beta}$ is
total field strength.

Removing $\partial_\sigma X^0$ from the first boundary condition, via
the third one, leads us
to define the variable $\tilde g_{\alpha \bar\beta}$ as
\bea
{\tilde g}_{\alpha \bar \beta} \equiv g_{\alpha \bar \beta}
- g_{\alpha 0} \frac {a_{\bar \beta}}{a_0},
\eea
where ${\bar \beta} \in \{1, 2, \cdot \cdot \cdot,p\}$.
To obtain desirable boundary conditions
we demand ${\tilde g}_{0\bar \beta}$ to be zero which gives
$a_{\bar \beta} = \frac {a_0}{g_{00}} g_{0 \bar \beta}$.
In addition, we impose the extra conditions
${\cal{F}}_{\bar \alpha 0}=U_{0 \beta}=0$.
Therefore, the boundary condition of open string along the
brane directions takes the form
\bea
({\tilde g}_{\bar \alpha {\bar \beta}}\partial_\sigma X^{\bar \beta} + 2\pi
i\alpha' {\cal{F}}_{\bar \alpha {\bar \beta}} \partial_\tau X^{\bar \beta}
+ U_{\bar \alpha {\bar \beta}} X^{\bar \beta})_{\sigma=0}=0.
\eea
The symmetric matrix ${\tilde g}_{\bar \alpha {\bar \beta}}=
g_{{\bar \alpha}{\bar \beta}}- \frac {g_{00}}{{a_0}^2}
a_{\bar \alpha} a_{\bar \beta}$
effectively possesses the treatment of a metric in the brane volume.
By having the coordinates $\{X^{\bar \alpha}\}$,
we are able to specify $X^0$ via its equation
$\partial^2 X^0 =0$ and the boundary condition:
$\partial_\sigma X^0|_{\sigma=0}=-\frac {a_{\bar \alpha}}{a_0}
\partial_\sigma X^{\bar \alpha}|_{\sigma=0}$.
\section{Non-commutativity variables}

The open string propagator ${\cal{G}}^{{\bar \alpha}{\bar \beta}}$ can be
calculated via the equation
\bea
\partial {\bar \partial}{\cal{G}}^{{\bar \alpha}{\bar \beta}}(z,z') = -2\pi
{\tilde g}^{{\bar \alpha}{\bar \beta}}
\delta^{(2)}(z-z'),
\eea
and the boundary condition
\bea
[(\tilde g + 2\pi \alpha' {\cal{F}})\partial {\cal{G}}- (\tilde g -2\pi
\alpha'{\cal{F}}){\bar \partial}{\cal{G}}
-i{U}{\cal{G}}]^{{\bar \alpha}{\bar \beta}}|_{\sigma=0}=0,
\eea
where the complex variable is $z=\tau + i\sigma$.
The solution of these equations is given by
\bea
{\cal{G}}^{{\bar \alpha}{\bar \beta}}(z,z') = &-&\alpha' \tilde
g^{{\bar \alpha}{\bar \beta}} \ln |z-z'|
\nonumber\\
&+& \frac {\alpha'}{2} \sum ^\infty _{n=1} {\bigg{(}
\frac {\tilde g - 2\pi \alpha' {\cal {F}} - \frac
{\alpha'}{2n} U} {\tilde g + 2\pi \alpha'{\cal
{F}} + \frac {\alpha'}{2n} U}
\bigg{)}}^ {({\bar \alpha}{\bar \beta})}
\frac{(z{\bar z'}) ^ n + ({\bar z}z')^n}{n}
\nonumber\\
&+& \frac {\alpha'}{2} \sum ^\infty _{n=1} {\bigg{(}
\frac {\tilde g - 2\pi \alpha' {\cal {F}} - \frac{\alpha'}{2n}
U} {\tilde g + 2\pi \alpha'{\cal{F}} + \frac {\alpha'}{2n} U}
\bigg{)}}^{[{\bar \alpha}{\bar \beta}]}
\frac{(z{\bar z'})^n - ({\bar z}z')^ n}{in}.
\eea

According to the prototype initiated in the Ref. \cite{3},
the above propagator defines an open string
metric and a non-commutativity parameter, for each string mode,
as in the following
\bea
G^{{\bar \alpha}{\bar \beta}}_n &=& \bigg{(} \frac{1}{\tilde g
+ 2\pi \alpha' {\cal{F}}
+ \frac {\alpha'}{2n} U} \bigg{)}^{({\bar \alpha}{\bar \beta})}
\nonumber\\
&=&\bigg{(} \frac{1}{{\tilde g} + 2\pi \alpha' {\cal{F}}+
\frac{\alpha'}{2n} U} \;\; ({\tilde g}+\frac {\alpha'}{2n} U)
\;\; \frac {1}{{\tilde g}
- 2\pi \alpha' {\cal {F}}+\frac {\alpha'}{2n} U}
\bigg{)}^{{\bar \alpha}{\bar \beta}},
\eea
\bea
\theta^{{\bar \alpha}{\bar \beta}}_n &=&
2\pi \alpha' \bigg{(} \frac{1}{\tilde g + 2\pi
\alpha' {\cal {F}} + \frac {\alpha'}{2n} U}
\bigg{)}^{[{\bar \alpha}{\bar \beta}]}
\nonumber\\
&=& -(2\pi \alpha')^2 \bigg{(} \frac{1}{{\tilde g}
+ 2\pi \alpha' {\cal {F}}+\frac {\alpha'}{2n} U}
\;\;{\cal{F}} \;\; \frac{1}{{\tilde g} - 2\pi \alpha' {\cal{F}}
+\frac{\alpha'}{2n} U} \bigg{)}^{{\bar \alpha}{\bar \beta}},
\eea
where $n \in \{1,2,3,\cdot\cdot\cdot\}$.
Note that these variables depend on the positive string
mode numbers. Analog results can be seen for noncommutative
$D$-branes in the pp-wave background \cite{17}.
The peculiar result that we obtained is a consequence
of the tachyon field in the volume of the brane. In the
absence of the tachyon all modes of the open string probe
the same value for each of these variables. However, the
tachyon splits this degeneracy. By adjusting the parameters
${\tilde g}_{{\bar \alpha}{\bar \beta}}$,
${\cal{F}}_{{\bar \alpha}{\bar \beta}}$ and $U_{{\bar \alpha}{\bar \beta}}$
one can receive expedient values of the non-commutativity variables.

Now we introduce the averaged values of the non-commutativity
variables
\bea
&~& {\bar G}^{{\bar \alpha}{\bar \beta}}=
\lim_{N\rightarrow\infty} {\frac{1}{N} \sum^N_{n=1}
G^{{\bar \alpha}{\bar \beta}}_n},
\nonumber\\
&~& {\bar \theta}^{{\bar \alpha}{\bar \beta}}= \lim_{N\rightarrow
\infty} {\frac{1}{N} \sum^N_{n=1} \theta^{{\bar \alpha}{\bar
\beta}}_n}. \eea By using the Eqs. (10) and (11), these variables
are acquired as \bea &~& {\bar G}^{{\bar \alpha}{\bar \beta}}=
\bigg{(} \frac {1}{\tilde g + 2\pi \alpha' {\cal {F}}} \;\;
\tilde g \;\; \frac {1}{\tilde g - 2\pi \alpha' {\cal {F}}}
\bigg{)}^{{\bar \alpha}{\bar \beta}},
\nonumber\\
&~& {\bar \theta} ^{{\bar \alpha} {\bar \beta}} = -(2\pi
\alpha')^2 \bigg{(} \frac {1}{\tilde g + 2\pi \alpha' {\cal {F}}}
\;\;{\cal{F}} \;\; \frac {1}{\tilde g - 2\pi \alpha' {\cal{F}}}
\bigg{)}^{{\bar \alpha}{\bar \beta}}. \eea These are independent
of the tachyon matrix. Therefore, they are counterparts of the
well-known standard non-commutativity variables, as expected. In
fact, due to the dilaton parameters $\{a_{\bar \alpha}|{\bar
\alpha} = 1, \cdot\cdot\cdot, p\}$ in the matrix ${\tilde
g}_{{\bar \alpha} {\bar \beta}}$, these are generalized version
of the ordinary case.
\subsection{Non-commutativity after tachyon condensation}

Ground state of the bosonic open string has negative
mass squared and hence is tachyonic.
In fact, the tachyon field describes the dynamics of an unstable
D$p$-brane. That is, after tachyon condensation (i.e. when at
least one of the nonzero elements of the matrix $U_{{\bar \alpha}
{\bar \beta}}$ goes to infinity) the brane becomes unstable.
Therefore, a D$p$-brane in the presence of a tachyon field
decays to the lower dimensional branes or decays to the
closed string vacuum.

According to the Eqs. (10) and (11), since each mode feels its
own non-commutativity, tachyon condensation depends on the mode
number. Let tachyon condensation to take place in the
$x^p$-direction of the brane. For the finite mode numbers this
implies the limit $\frac{1}{n}U_{pp}\rightarrow \infty$. Thus,
the last columns and the lowest rows of the matrices $G_n$ and
$\theta_n$ vanish. The nonzero $(p-1)\times (p-1)$ matrices
inside the matrices $G_n$ and $\theta_n$ elaborate the open
string metric and non-commutativity parameter of a D$(p-1)$-brane.
In other words, the D$p$-brane loses its extension along the
direction $x^p$, as expected. The features of the new
non-commutativity variables are similar to the previous case,
i.e. the Eqs. (10) and (11), with ${\bar \alpha},{\bar \beta} \in
\{1,2,\cdot \cdot \cdot,p-1\}$. For very large mode numbers if
the quantity $\frac{1}{n}U_{pp}$ tends to infinity again we have
the above discussion. If the infinite values of $U_{pp}$ and $n$
lead to a finite value for $\frac{1}{n}U_{pp}$, then tachyon
condensation does not occurs.

We observe that the averaged values of the non-commutativity
variables are independent of the tachyon, and hence they
are not affected by condensation of the tachyon.
In the Sec. 5 the tachyon condensation again will be illustrated.
\section{Scaling limits}

\subsection{The zero slope limit}

Now we impose the zero slope limit ($\alpha'\rightarrow 0$) on the open
string variables. This is useful for studying the low energy
behavior of the open string. Since open strings are sensitive
to $G$ and $\theta$, we should take the limit in the manner
that these variables to be fixed rather than the closed
string parameters \cite{3}. Therefore, we apply the limits
$\alpha' \sim \epsilon^\frac{1}{2} \rightarrow 0$,
$g_{{\bar \alpha}{\bar \beta}} \sim \epsilon \rightarrow 0$ and
$U_{{\bar \alpha}{\bar \beta}}\sim \epsilon^{\frac{1}{2}}$,
while ${\cal{F}}$ is fixed. The Eq. (5) gives the scaling
$\tilde g_{{\bar \alpha}{\bar \beta}} \sim \epsilon \rightarrow 0$.
In these limits we receive the following quantities
\bea
&~& G^{{\bar \alpha}{\bar \beta}}_n =
-\frac{1}{(2\pi \alpha')^2}\bigg {(}
{\cal{F}}^{-1}\;({\tilde g}+\frac{\alpha'}{2n}U)\;{\cal{F}}^{-1}
\bigg{)}^{{\bar \alpha}{\bar \beta}},
\nonumber\\
&~& \theta^{{\bar \alpha}{\bar \beta}}_n
= ({\cal{F}}^{-1})^{{\bar \alpha}{\bar \beta}},
\eea
\bea
&~& {\bar G}^{{\bar \alpha}{\bar \beta}} =
-\frac{1}{(2\pi\alpha')^2}\bigg{(}
{\cal{F}}^{-1}\;{\tilde g}\;{\cal{F}}^{-1}
\bigg{)}^{{\bar \alpha}{\bar \beta}},
\nonumber\\
&~& {\bar \theta}^{{\bar \alpha}{\bar \beta}} =
({\cal {F}}^{-1})^{{\bar \alpha}{\bar \beta}}.
\eea
Since ${\tilde g}_{{\bar \alpha}{\bar \beta}}$,
$\alpha'U$ and $\alpha'^2$ have the same limiting
behavior the above variables are finite.
\subsection{Small ${\cal{F}}$ and $U$ limit}

A small tachyon and total gauge field limit can be obtained by using
the scaling $g_{{\bar \alpha}{\bar \beta}} \sim \epsilon$,
$U_{{\bar \alpha}{\bar \beta}}\sim \epsilon$ and
${\cal{F}}_{{\bar \alpha}{\bar \beta}}\sim \epsilon^\frac{1}{2}$.
In this limit the corresponding non-commutativity variables
reduce to
\bea
&~& G^{{\bar \alpha}{\bar \beta}}_n =
-\frac{1}{(2\pi \alpha')^2}\bigg {(}
{\cal{F}}^{-1}\;({\tilde g}+\frac{\alpha'}{2n}U)\;{\cal{F}}^{-1}
\bigg{)}^{{\bar \alpha}{\bar \beta}},
\nonumber\\
&~& \theta^{{\bar \alpha}{\bar \beta}}_n
= ({\cal{F}}^{-1})^{{\bar \alpha}{\bar \beta}},
\eea
\bea
&~& {\bar G}^{{\bar \alpha}{\bar \beta}} =
-\frac{1}{(2\pi\alpha')^2}\bigg{(}
{\cal{F}}^{-1}\;{\tilde g}\;{\cal{F}}^{-1}
\bigg{)}^{{\bar \alpha}{\bar \beta}},
\nonumber\\
&~& {\bar \theta}^{{\bar \alpha}{\bar \beta}} =
({\cal{F}}^{-1})^{{\bar \alpha}{\bar \beta}}.
\eea
Though the features of these variables and their counterparts
in the zero slope limit are the same but they have different
values. For example, the non-commutativity parameters
in the Eqs. (14) and (15) are finite while in the Eqs. (16)
and (17) they are very large. However, the closed string
metrics in both cases are finite.
\subsection{Large ${\cal{F}}$ and $U$ limit}

Now we examine large values for the matrices ${\cal{F}}$ and $U$,
\bea
\alpha' \sim \epsilon^\frac{1}{2},\;\;\;\;
{{\cal{F}}}_{{\bar \alpha}{\bar \beta}}\sim \epsilon^{-\frac{1}{2}},\;\;\;\;
U_{{\bar \alpha}{\bar \beta}}\sim \epsilon^{-\frac{1}{2}},\;\;\;\;
g_{{\bar \alpha}{\bar \beta}} \sim \epsilon \rightarrow 0.
\eea
This scaling leads to the following equations
\bea
&~& G^{{\bar \alpha}{\bar \beta}}_n =
\frac{2n}{\alpha'}\bigg {(}\frac{1}{U+4\pi n{\cal{F}}}\;U\;
\frac{1}{U-4\pi n{\cal{F}}}\bigg {)}^{{\bar \alpha}{\bar \beta}},
\nonumber\\
&~& \theta^{{\bar \alpha}{\bar \beta}}_n=-(4\pi n)^2\bigg {(}
\frac{1}{U+4\pi n{\cal{F}}}\;{\cal{F}}\;
\frac{1}{U-4\pi n{\cal{F}}}\bigg {)}^{{\bar \alpha}{\bar \beta}}.
\eea
\bea
&~& {\bar G}^{{\bar \alpha}{\bar \beta}} =
-\frac{1}{(2\pi\alpha')^2}\bigg{(}
{\cal{F}}^{-1}\;{\tilde g}\;{\cal{F}}^{-1}
\bigg{)}^{{\bar \alpha}{\bar \beta}},
\nonumber\\
&~& {\bar \theta}^{{\bar \alpha}{\bar \beta}} =
({\cal {F}}^{-1})^{{\bar \alpha}{\bar \beta}}.
\eea
From the open string metrics only $G_n$ is finite while
${\bar G}$ behaves like $\epsilon$. The non-commutativity
parameters $\theta_n$ and ${\bar \theta}$ also go to
zero similar to $\epsilon^\frac{1}{2}$.
\subsection{Small tachyon approximation}

An interesting approximation is given by the small tachyon
matrix elements. In this case the non-commutativity variables
are truncated to the following versions
\bea
&~& \lim_{U \rightarrow 0} G^{{\bar \alpha}{\bar \beta}}_n=
{\bar G}^{{\bar \alpha}{\bar \beta}}
+\frac{\alpha'}{2n}\bigg{(}
\frac{1}{{\tilde g} + 2\pi \alpha'{\cal{F}}}\;U\;
\frac{1}{{\tilde g} - 2\pi \alpha'{\cal{F}}}
\nonumber\\
&~& -U\;\frac{1}{{\tilde g} + 2\pi \alpha'{\cal{F}}}\;{\bar G}
-{\bar G}\;\frac{1}{{\tilde g} - 2\pi \alpha'{\cal{F}}}\;U
\bigg{)}^{{\bar \alpha}{\bar \beta}}
+{\cal{O}}(U^2)^{{\bar \alpha}{\bar \beta}},
\eea
\bea
\lim_{U \rightarrow 0}\theta^{{\bar \alpha}{\bar \beta}}_n=
{\bar \theta}^{{\bar \alpha}{\bar \beta}}
-\frac{\alpha'}{2n}\bigg{(} {\bar \theta}\;
\frac{1}{{\tilde g} - 2\pi \alpha'{\cal{F}}}\;U
+U\;\frac{1}{{\tilde g} + 2\pi \alpha'{\cal{F}}}\;
{\bar \theta}\bigg{)}^{{\bar \alpha}{\bar \beta}}
+{\cal{O}}(U^2)^{{\bar \alpha}{\bar \beta}}.
\eea
We observe that the first terms of these limits are the averaged
non-commutativity variables, as expected. The first order
corrections also slow down by the factor $1/n$.
\section{Examples}

To have more physical intuitive of the above general set-up,
we describe the non-commutativity variables
for two simple examples. These are D1-brane and D2-brane.

\subsection{D1-Brane}

Consider a D1-brane lying in the $X^1$-direction. The general
forms of the closed string metric and the matrix $U$ can be
written as
\bea
g_{\alpha \beta} =
\left(\begin{array}{cc}
-g_0 &  g_1  \\
g_1 & g_2 \\
\end{array} \right),
\;\;\;\;\;
\;{U}_{\alpha \beta} = \left(\begin{array}{cc}
0 &  0\\
0 & u_2 \\
\end{array} \right).
\eea
Note that in our formulation there are
${\cal {F}}_{\bar \alpha 0} = U_{0 \beta} = 0$.
According to ${\tilde g}_{11}=g_2 + \frac {{g_1}^2}{g_0}$,
the open string metrics reduce to
\bea
&~& G_{(n)11}= g_2 + \frac {{g_1}^2}{g_0}+ \frac {\alpha'}{2n} u_2 ,
\nonumber\\
&~& {\bar G}_{11}= g_2 + \frac {{g_1}^2}{g_0}.
\eea
Since there is only one spatial direction the non-commutativity
parameters vanish.

\subsection{D2-Brane}

For a D2-brane in the $X^1 X^2$-plane there are
\bea
g_{\alpha \beta} =
\left(\begin{array}{ccc}
-g_0 &  g_1 & g_3 \\
g_1 & g_2 & g_4\\
g_3 & g_4 & g_5 \\
\end{array} \right),\;\;\; {\cal {F}}_{\alpha \beta}= \left(\begin{array}{ccc}
0 &  0 & 0 \\
0 & 0 & b\\
0 & -b & 0 \\
\end{array} \right),\;\;\; U_{\alpha \beta}= \left(\begin{array}{ccc}
0 &  0 & 0 \\
0 & u_2 & u_4\\
0 & u_4 & u_5 \\
\end{array} \right).
\eea
The closed string metric defines the matrix $\tilde g$ as in the following
\bea
\tilde g_{{\bar \alpha} {\bar \beta}} =
\left(\begin{array}{cc}
g_2 + \frac {{g_1}^2}{g_0} &  g_4 + \frac {g_1 g_3 }{g_0} \\
g_4 + \frac {g_1 g_3}{g_0} & g_5 + \frac {{g_3}^2}{g_0} \\
\end{array} \right)\;,\;\;\;\;\; {\bar \alpha}, {\bar \beta} \in \{1,2\}.
\eea
Therefore, the non-commutativity variables take the forms
\bea
&~& G^{{\bar \alpha}{\bar \beta}}_n =
\left(\begin{array}{cc}
\frac {\frac {g_3^2}{g_0} + g_5 + \frac{\alpha'}{2n} u_5}{P_n} &
-\frac {2 n (2n g_1 g_3  + 2n g_0 g_4 + \alpha g_0 u_4)}{Q_n} \\
-\frac {2 n (2n g_1 g_3  + 2n g_0 g_4 + \alpha' g_0 u_4 )}{Q_n} &
\frac {\frac {g_1^2}{g_0} + g_2 + \frac {\alpha'} {2n} u_2}{P_n}\\
\end{array} \right),
\nonumber\\
&~& \theta^{{\bar \alpha} {\bar \beta}}_n= \left(\begin{array}{cc}
0 & -\frac {8  \pi \alpha' n^2 b g_0}{Q_n} \\
\frac {8  \pi \alpha' n^2 b g_0}{Q_n}& 0 \\
\end{array} \right),
\eea
where the variables $P_n$ and $Q_n$ have the following definitions
\bea
P_n &=& \bigg{(} \frac{g_1 g_3}{g_0} + g_4 - 2 \pi \alpha' b +
\frac {\alpha'}{2n} u_4 \bigg{)} \bigg{(} \frac {g_1 g_3}{g_0} + g_4 +
2 \pi \alpha' b + \frac {\alpha'}{2n} u_4 \bigg{)}
\nonumber\\
&+& \bigg{(} \frac {g_1^2}{g_0}
+ g_2 + \frac {\alpha' }{2 n}u_2 \bigg{)} \bigg{(} \frac {g_3^2}{g_0}+ g_5 +
\frac {\alpha'}{2 n} u_5 \bigg{)},
\nonumber\\
Q_n &=& 4 n^2 g_0 {g_4}^2 + 2n \alpha' g_3^2 u_2 + 2n \alpha' g_0 g_5 u_2
\nonumber\\
&-& 4n \alpha' g_0 g_4 u_4 + 16 \pi^2 n^2 \alpha'^2 b^2 g_0 - \alpha'^2 g_0 u_4^2
\nonumber\\
&+& \alpha'^2 g_0 u_2 u_5 - 4n g_1 g_3(2n g_4 + \alpha' u_4 )+
2n g_1^2 (2 n g_5 + \alpha' u_5 )
\nonumber\\
&+& 2n g_2 (2n g_3^2 + 2n g_0 g_5 + \alpha' g_0 u_5).
\eea

Regarding to the tachyon condensation consider the limit
$(U_{22}=u_5) \rightarrow \infty$. The equations (27) implies
that the non-commutativity matrix vanishes and the only nonzero
element of the open string metric reduces to $G^{11}_n=1/(g_2 +
\frac {{g_1}^2}{g_0}+ \frac {\alpha'}{2n} u_2)$. This is exactly
the inverse of the first equation in (24). That is, the D2-brane
has deformed to a D1-brane, as expected.
\section{Conclusions and summary}

The non-commutativity of open strings, which are attached to a
D$p$-brane in the presence of the massless fields:
Kalb-Ramond, a $U(1)$ gauge field, dilaton and a tachyon
field which is massive, was investigated. Presence of the linear
dilaton field effectively deforms the closed string metric,
and hence the non-commutativity variables. Appearance of
various parameters due to the background fields, i.e.
$\{B_{{\bar \alpha}{\bar \beta}},\;F_{{\bar \alpha}{\bar \beta}},\;
a_{\bar \alpha},\;U_{{\bar \alpha}{\bar \beta}}\}$
enables us to adjust these variables to desirable values.

The non-commutativity variables are influenced by
all background fields, but the tachyon field has more control
on them. Precisely, the tachyon field decomposes a
non-commutativity which has been originated by the Kalb-Ramond
and the $U(1)$ gauge fields into infinite number of
non-commutativities. That is, each open string mode feels its own
non-commutativity. The average values of the non-commutativity
variables are independent of the tachyon.
Accurately they are equal to the
non-commutativity of infinite massive states of open string, or
equivalently they are the ordinary non-commutativity variables, which
appear in the systems without tachyon field.
As expected, tachyon condensation deform the non-commutativity
variables of a D$p$-brane to that of a D$(p-1)$-brane.
However, this reduction for light string modes always takes place,
while for very large mode numbers its occurence depends on the
orders of the infinities of ``$n$'' and ``$U_{pp}$''.

Finally, various scaling limits were studied. The tachyon matrix
enabled us to introduce new appealing scaling limits.

Note that for a moving D$p$-brane in our set-up,
with velocity perpendicular to the brane directions, the
non-commutativity structure is the same as we studied.
By contrast, a moving D$p$-brane along itself generates
a new non-commutativity structure which can be investigated.

\end{document}